\documentclass[a4paper,fleqn]{cas-sc}

\usepackage[authoryear]{natbib}
\usepackage{lineno}

\newcommand\apjl{{ApJL}}     
\newcommand\apjs{{ApJS}}
\newcommand\ao{{ApOpt}}
\newcommand\aap{{A\&A}}

\newcommand{\edits}[1]{\textcolor{black}{#1}}
\def\tsc#1{\csdef{#1}{\textsc{\lowercase{#1}}\xspace}}
\tsc{WGM}
\tsc{QE}

\begin{document}
\let\WriteBookmarks\relax
\def\floatpagepagefraction{1}
\def\textpagefraction{.001}

\shorttitle{Dynamical Feasibility of (3) Juno as a Parent Body of the H chondrites}    

\shortauthors{Noonan et al. }  

\title [mode = title]{Dynamical Feasibility of (3) Juno as a Parent Body of the H Chondrites}  



%

\author[1,2]{John W. Noonan}[orcid=0000-0003-2152-6987]

\cormark[1]


\ead{noonan@auburn.edu}

\ead[url]{jnoons.squarespace.com}


\affiliation[1]{organization={Lunar and Planetary Laboratory,University of Arizona},
            addressline={1629 E University Blvd}, 
            city={Tucson},
            postcode={85721-0092}, 
            state={AZ},
            country={USA}}
\affiliation[2]{organization={Department of Physics, Auburn University},
            addressline={Edmund C. Leach Science Center}, 
            city={Auburn},
            postcode={36849}, 
            state={AL},
            country={USA}}
\affiliation[3]{organization={Planetary Science Institute},
            addressline={1700 East Fort Lowell, Suite 106}, 
            city={Tucson},
            postcode={85719}, 
            state={AZ},
            country={USA}}
\affiliation[4]{organization={Planetary Science Directorate, Southwest Research Institute},
            addressline={Suite 300, 1050 Walnut St}, 
            city={Boulder},
            postcode={80302}, 
            state={CO},
            country={USA}}

\author[1,3]{Kathryn Volk}[orcid=0000-0001-8736-236X]
\author[4]{David Nesvorn{\'y}}[]
\author[4]{William F. Bottke}[]



\cortext[1]{John W. Noonan}



\begin{abstract}
We test the hypothesis that (3) Juno is a parent body of the H chondrites with dynamical modeling of an asteroid-family-forming impact and comparison to current observational data. Using a dynamical model that includes the Yarkovsky force on a simulated Juno family and a simplified cosmic ray exposure age model we examine the expected distribution of Juno family members in both the main belt and near-Earth orbits over 300 Myrs and the cosmic ray exposure distribution for fragments exiting the main belt via the 3:1J, 5:2J, and 8:3J mean motion resonances. We find that the smallest modeled ($D<$10 m) family members of \edits{(3) Juno} cannot be directly responsible for the observed H chondrite flux and that the breakup of larger family members creates an CRE distribution that resembles the measured H chondrite CRE distribution but is still unable to adequately explain the significant number of H chondrites with CRE ages of 6-8 Myrs. A similar model was performed for the asteroid (6) Hebe, another parent body candidate, and produced a CRE age distribution that is inconsistent with the measured H chondrite CRE ages. We also find from our dynamical models that we can expect $<$7 km-scale Juno family members in near-Earth orbits in the present day, consistent with the recent discovery of the shock-darkened H chondrite-like asteroid (52768) 1998 \edits{OR$_{2}$}. 
\end{abstract}


\begin{highlights}
\item We use dynamical modeling with the Rebound package to evaluate the ability of (3) Juno to track potential H chondrite family members for 300 Myrs after a family forming event and the TrackMet cosmic ray exposure age model to produce a synthetic cosmic ray exposure age distribution for both \edits{(3) Juno} and (6) Hebe from the collisional cascade of family members. 
\item We find that the resulting orbital distribution of family members from a Juno family-forming event 300 Myrs \edits{ago} is consistent with both the observed number of Juno family members near the 8:3J and the recent discovery of a shock-darkened H chondrite-like Near Earth Object. 
\item We find that the modeled cosmic ray exposure age distribution for Earth-impacting fragments from a Juno family\edits{-}forming event is consistent with being drawn from the same population as the measured H chondrite cosmic ray exposure age distribution from meteorites. Fragments evolving from a similar family\edits{-}forming event 300 Myrs ago \edits{(6) Hebe} are statistically likely (p$<$0.01) to be drawn from a different distribution.
\end{highlights}

\begin{keywords}
\sep Asteroids, dynamics \sep Resonances, orbital \sep Meteorites
\end{keywords}

\maketitle

\section{Introduction}
Identifying the source regions of meteorites and Near-Earth \edits{Objects (NEOs)} serves to improve our understanding of early solar system formation by providing traceability. By determining where the meteorites originated and how they formed we can place the structural and chemical observations of meteorites into the broader context necessary to improve the timeline of the solar system's early days \citep{2006AREPS..34..157B,2017DPS....4910003B}. Of all meteorites cataloged, 86\% are classified as ordinary chondrites, and approximately 34\% are H chondrites 
\citep[Meteoritical Catalog,][]{burbine_meteoritic_2002}.
H chondrites are distinct from the other large ordinary chondrite types L and LL based on their iron content (H for high, L for low, LL for very low; \citealt{burbine_meteoritic_2002}). 
With H chondrites representing such a large portion of meteorites for direct laboratory experiments, it is crucial to understand the dynamical feasibility of different parent bodies as their potential source.

We have some constraints on candidates for the H chondrite parent body.
Early work showed spectral similarities between H chondrites and the S type asteroid (6) Hebe \citep{gaffey_mineralogical_1993,1997M&PS...32..903M, 1998M&PS...33.1281G}, but without a confirmed asteroid family to derive from, this link is tenuous \citep{2015PDSS..234.....N}. The criteria that a parent body of the H chondrites must fulfill are more completely discussed in \cite{2019AJ....158..213N}, but we will briefly outline them here.

\textit{Thermal modeling constraints}: In an effort to constrain the size of the H chondrite parent body, thermal models based on the meteorites themselves, specifically the thermal alteration, 
have been implemented \citep{2003Natur.422..502T,2003M&PS...38..711G,2005GeCoA..69..505A,2007M&PS...42.1337B,2008E&PSL.270..106K,2010GeCoA..74.5410H,2012A&A...545A.135H,2013GeCoA.105..206G,2013GeCoA.119..302M,2014GeCoA.136...13S,2017GeCoA.200..201B}. These studies focus on the amount of $^{26}$Al heating the parent body experienced and the resulting levels of differentiation; these can be affected by the meteorite samples chosen for analysis.
The majority of these efforts suggest that the H chondrite parent body is larger than 200 km in diameter, with the most recent study \citep{2017GeCoA.200..201B} suggesting that it was larger than 275 km. 

\textit{Cosmic ray exposure ages}: Of all studied H chondrites, nearly half exhibit cosmic ray exposure ages between 7 and 8 million years, a trait unique to this meteorite class \citep{2006mess.book..829E}. Such a specific age leads us to believe that the H chondrites have one dominant parent body or collisional family, which suggests two possibilities: the parent body was located near an efficient mean motion resonance (MMR) for fast delivery of a resulting family to Earth-crossing orbits, or the body itself was near or on an Earth-crossing orbit when the breakup event happened. Both cases could result in nearly half of the H chondrites having such a distinct cosmic ray exposure age.

\textit{$^{40}$Ar-$^{39}$Ar Shock degassing ages}: Studies of argon trapped in meteorites are useful for determining dates for significant impact events on large asteroids, specifically for impacts with resulting craters larger than 10 km in diameter. The heating from these events is large enough to go beyond the loss threshold of argon, providing a useful clue about a meteorite's parent body's impact history. For H chondrites, the $^{40}$Ar-$^{39}$Ar shock degassing ages show events at 4.4-4.5, 3.5-4.1, and 0-1 billion years ago, but the largest density of ages is found at 300 million years ago \citep{2014GSLSP.378..333S}. This shock degassing age is in rough agreement with the estimated age of the Juno family \citep{2015PDSS..234.....N,2019AJ....158..213N}. 

\textit{Paleomagnetic evidence}: The metallic melt impact breccia Portales Valley, an H6 meteorite \citep{2018GeCoA.234..115R}, provides evidence for a magnetic field in parent body $\sim$4.5 billion years ago \citep{2014GSLSP.378..333S,2016AGUFM.P53D..02B}. Accretion and dynamo modeling of the parent body size to achieve the measured 10 $\mu$T magnetic field in the sample shows that diameters between 230 and 320 km are necessary. This is consistent with the size estimates from the thermal modeling constraints.

\begin{figure}
    \centering
    \includegraphics[width=0.7\linewidth]{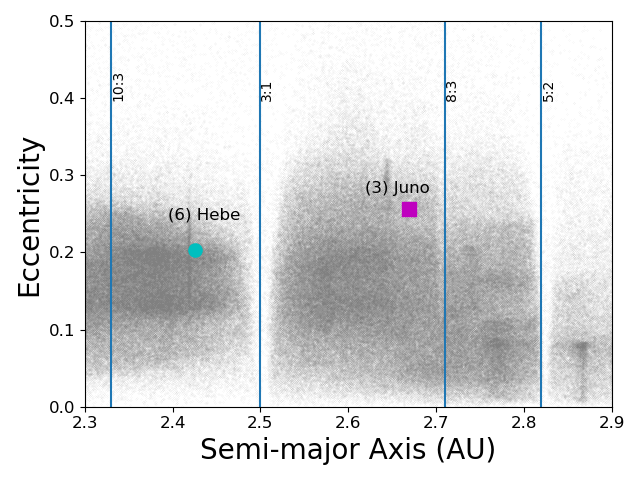}
    \caption{Plot of eccentricity vs. semi-major axis depicting (3) Juno and (6) Hebe relative to major resonances. Other main belt asteroids are plotted in grey. (6) Hebe is situated closer to the 3:1 resonance with Jupiter, but lacks a family capable of sustaining delivery of asteroids to the resonance for injection into near-Earth orbits. (3) Juno has a family, but is farther away from the strong resonance.}
    \label{fig:juno_and_Hebe}
\end{figure}

As pointed out in \cite{2019AJ....158..213N}, all of these criteria make the asteroid (3) Juno an excellent candidate for a parent body of the H chondrites. In addition, the \edits{S type } surface of \edits{(3) Juno} \citep{demeo_extension_2009,gaffey_mineralogical_1993} has an 89\% chance of being H chondrite-like based on a Bayesian classifier, which compares band area ratios and olivine and pyroxene abundances derived from the near-infrared spectrum of asteroids to meteoritic samples \citep{2019AJ....158..213N}. A similar spectral relationship has been reported between H chondrites and the asteroid (6) Hebe \citep{farinella_injection_1993,1997M&PS...32..903M,gaffey_asteroid_1998,fieber2020family}, which has more favorable access to the $\nu$6 and 3:1 MMR with Jupiter (hereafter we will use `J' to indicate Jupiter's MMRs, e.g., 3:1J). Thus, the difficulty lies in the delivery of Juno family members to Earth-crossing orbits (Figure \ref{fig:juno_and_Hebe}). \edits{(3) Juno} sits nearly halfway between the 3:1J and 5:2J MMRs, and quite close to the 8:3J MMR. The efficiencies of these resonances at producing Earth-crossing objects (NEOs) is contingent on many factors and needs to be numerically modeled \citep{1997Sci...277..197G,1998M&PS...33..999M}.  \edits{(3) Juno} is not particularly well poised to create family members that quickly reach the strong 3:1J resonance, unlike another candidate (6) Hebe \citep{gaffey_asteroid_1998,fieber2020family}. The weaker 8:3J resonance is much closer (cf. Figure \ref{fig:juno_and_Hebe}), but \edits{is not a clear source to} deliver Juno family members; its scattering efficiency has only been characterized for objects with the orbital inclinations of (1) Ceres and (2) Pallas ($10.6^\circ$ and $34.9^\circ$, respectively; \citealt{de2010origin,todorovic2018dynamical,2022MNRAS.509.3842K}). 

\edits{However, (3) Juno has a large associated family that could be contributing. The Juno family consists of up to 1683 known members \citep{nesvorny_nesvorny_2015}, with \cite{Knezevic:2014} finding a slightly different number of associated family members,  1691 . The family's most recent age has been determined to be $\sim$300 MYrs \citep{2019AJ....158..213N}, while \cite{2015Icar..257..275S} finds discordant ages for the inner and outer slopes of the family that places the age of the Juno family between 370 and 550 Myrs, with a standard deviation of $\sim$160 Myrs. A cross check between Juno family members in the AstDyS family list \citep{Knezevic:2003,Milani:2014,Knezevic:2014} and the NEOWISE albedo and diameter database of \cite{masiero_asteroid_2015} shows that 180 family members have pre-existing diameters and albedos measured in the near infrared. (3) Juno has an albedo of 0.238 measured by IRAS \citep{2015Icar..257..275S}, while NEOWISE reports an albedo of 0.214 \citep{masiero_asteroid_2015}. The similarity in age of the Juno family and the $^{40}$Ar-$^{39}$Ar shock degassing ages of H chondrites would seem to suggest a connection between the events, but the dynamical feasibility of (3) Juno as a major source of meteoritic material has never been investigated. }Understanding the dynamical process of delivering Juno family members to Earth-crossing orbits, especially \edits{in terms of} the efficiency and timescales, is of critical importance to understanding how reasonable \edits{(3) Juno} is as a parent body candidate. 

The discovery of the anisotropic re-radiation termed the Yarkovsky force and its effect on meteorite delivery to the Earth makes it clear that \edits{(3) Juno}'s candidicacy as an H chondrite parent body must be adequately explored to evaluate the dynamics and collisional evolution that shapes the delivery of Juno family members \edits{\citep{bottke_yarkovsky_2006,vokrouhlicky_efficient_2000}}. The Yarkovsky force results from the re-radiation of solar energy from an asteroid's surface, and produces a minute acceleration that is dependent on many properties: thermal emissivity, spin axis orientation, heliocentric distance, diameter, albedo, and so forth. The effect can happen on both diurnal and seasonal timescales, and over millenia can alter the semi-major axis of asteroids. For asteroids in the main belt this can bring them into unstable mean motion and secular resonances, scattering them into new orbits \edits{\citep{1997Sci...277..197G,morbidelli_orbital_1998,vokrouhlicky_efficient_2000}}.

In this paper we present two models of \edits{synthetic} Juno family members to address this problem. We implement a custom Yarkovsky force in the Rebound and ReboundX Python packages for dynamical integrations \citep{rein_rebound_2012,2020MNRAS.491.2885T}, as well as a collisional model to trace cosmic ray exposure ages, TrackMet, which has been previously used in \cite{nesvorny_asteroidal_2009} for the L chondrites. In Section \ref{sec:Methods} we describe the models and necessary assumptions. The results of the simulations are presented in Section \ref{sec:Results} and we discuss the implications in Section \ref{sec:Discussion}. A summary of the paper is provided in Section \ref{sec:Summary}.

\section{Methods}\label{sec:Methods}
Testing the viability of \edits{(3) Juno} as the parent body of the H chondrites requires two lines of inquiry. First, the \edits{synthetic} Juno family itself needs to be forward modeled to verify that the distribution of \edits{observed} family members, with a recent estimated age of 300 Myrs \citep{2019AJ....158..213N}, can be achieved by gravitational and non-gravitational effects with reasonable family member albedo and emissivity assumptions. Second, cosmic ray exposure modeling is needed to confirm that a 300 Myr year old asteroid family is capable of producing a measurable flux of meteorites with a cosmic ray exposure age distribution that is nearly entirely under 10 Myrs via subsequent collisions. We have chosen dynamical modeling with the Rebound \edits{\citep{2012A&A...537A.128R}} and ReboundX \edits{\citep{2020MNRAS.491.2885T}} Python packages to accomplish the former and the TrackMet model \edits{\citep{nesvorny_asteroidal_2009}} for the latter. 

\subsection{Rebound Modeling}
To perform the modeling of a synthetic Juno family, we used both the Rebound \citep{2012A&A...537A.128R} and ReboundX \citep{2020MNRAS.491.2885T} software packages. Integrations were performed within Rebound using both the IAS15 \citep{2015MNRAS.446.1424R} and WHFAST  \citep{2015MNRAS.452..376R} integrators to confirm particle behaviors were independent of integration method. Within each experimental model the Sun, Venus, Earth, Mars, Jupiter, Saturn, Uranus, and Neptune were all added as massive bodies via calls to JPL Horizons. The orbital elements for (3) Juno are also imported and used to populate the initial state of a \edits{synthetic} Juno family of asteroids. 

Both the diurnal and the seasonal Yarkovsky forces are calculated at each timestep for test particles according equations 4, 5 and 6 from \cite{bottke_yarkovsky_2006}. The force is used to then calculate updated acceleration and velocity vectors for the particle within Rebound. This was accomplished by creating a new force effect using the ReboundX framework \citep{2020MNRAS.491.2885T}, with the full Yarkovsky force calculation added as an extra force to the Rebound simulation. We note that this was done prior to the publication of \cite{ferich2022yarkovsky}, who performed a similar addition that is now freely available with the ReboundX source code. We encourage readers interested in pursuing Rebound modeling of the Yarkovsky force to explore their paper and software package.

To create the \edits{synthetic} Juno family asteroids, we randomly sample a position on a sphere representing \edits{(3) Juno}'s surface 1000 times and assign an initial velocity equal to \edits{(3) Juno}'s orbital velocity; we then add \edits{(3) Juno}'s escape velocity (120 m/s) to each particle in a direction normal to their position on the sphere (added as $dv_x$, $dv_y$, and $dv_z$ components). This is a simplified model for a collisional family, but adequately samples the resulting potential initial orbits of family members. This family is randomly generated for each run of the simulation, allowing a wide range of initial orbital parameters to be sampled.  Each family member is randomly assigned a diameter from a size frequency distribution with a power law index of -0.1. This distribution is not intended to match the size distribution of \edits{observed} Juno family members, but rather to adequately explore the orbital evolution of various sizes. Each modeled family member is given a rotational period consistent with its diameter in meters, a randomized axis of rotation relative to the solar radiation, an albedo the same as \edits{(3) Juno} (0.238, from IRAS and retrieved via\citealt{2015Icar..257..275S}), and an emissivity of 0.005 W/m$^{2}$, \edits{reasonable for somewhat dust-covered S type asteroids (i.e. not a bare rock surface) \citep{bottke_yarkovsky_2006}. The surface and bulk densities of the asteroids are assumed to be 1.7 and 2.5 g $cm^{3}$, respectively, as detailed in \cite{bottke_yarkovsky_2006}, which are slightly smaller than the bulk density of (3) Juno, ($\sim$3.3 g cm$^{-3}$,  \citealt{viikinkoski_vltsphere-_2015}).}.  Each of these parameters influences either the magnitude or effective direction of the Yarkovsky effect, and therefore must be assigned, or as in the case of rotation axis, randomly sampled. We note that these particle generation techniques are consistent between the Rebound and TrackMet codes to ensure comparability of results.

\begin{figure}
    \centering
    \includegraphics[width=0.7\linewidth]{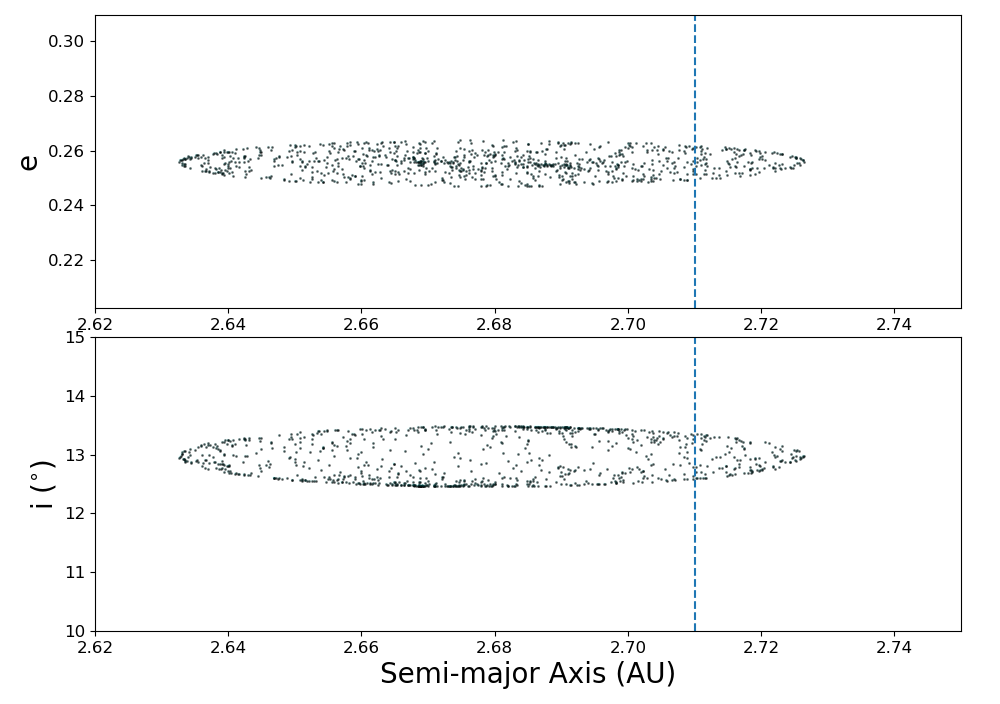}
    \caption{Initial semi-major axis, eccentricity, and inclination for randomly generated \edits{synthetic} Juno family members. The 8:3J resonance is marked with a dashed line. }
    \label{fig:init_orbital_elements}
\end{figure}

\begin{figure}
    \centering
    \includegraphics[width=0.7\linewidth]{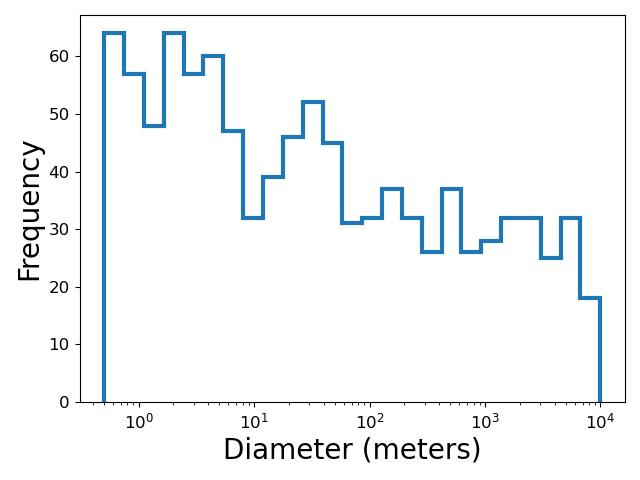}
    \caption{Histogram of randomly generated \edits{synthetic} Juno family member diameters used within the Rebound model. The diameters are randomly selected from a distribution with a power law of -0.1. }
    \label{fig:rebound_particle_sizes}
\end{figure}

Following generation of this family, each particle is integrated forward in time in a system containing the massless particle, a massless \edits{(3) Juno}, the terrestrial planets (except for Mercury), the gas giant planets, and the Sun. Each simulation included 100 randomly generated test particles of varying sizes, rotation rates, and initial ejection vectors from \edits{(3) Juno}. The initial semi-major axes ($a$), eccentricities ($e$), and inclinations ($i$) of these test particles are shown in Figure \ref{fig:init_orbital_elements}. A cumulative distribution plot of their diameters is shown in Figure \ref{fig:rebound_particle_sizes}. Ten simulations were run simultaneously on The University of Arizona Ocelote computing cluster for a total of 200 wall time hours each. Each simulation was given 320 Myrs of total integration time, with timesteps of 0.1 years. We did not implement any collisional cascade modeling for the Rebound model; the creation of such a large number of particles would have quickly overwhelmed the dynamical simulation, unlike the one dimensional TrackMet model described in Section~\ref{sec:trackmet}. 
To reduce computational time, we also added a conditional test particle removal check that identified test particles that reached orbits that could have close encounters with Mars or Jupiter. The conditions for this removal process are $a(1-e) \leq 1.7$~au and  $a(1+e) \geq 4.1$~au , where 1.7 and 4.1 au represent the semi-major axis plus/minus three times the size of the Hill radius for Mars and Jupiter, respectively. 

\subsection{TrackMet Modeling}\label{sec:trackmet}
To explore the collisional evolution and cosmic ray exposure ages of \edits{synthetic} Juno family members, we implemented the TrackMet model described in \cite{nesvorny_asteroidal_2009}. The model tracks the collisions for randomly generated individual family members as they evolve via the Yarkovsky effect, implemented identically to the Rebound version with the exception of timestep size, but does not execute a full dynamical model; the model only tracks radial progression in time steps of 500,000 years.  Particles are collisionally evolved based on their diameters according to the rates described in \cite{bottke_linking_2005}, as in \cite{nesvorny_asteroidal_2009}. When collisions occur, a new distribution of particles is created, conserving the mass of the original, and the model continues to track all new particles in their radial evolution. When particles evolve to sizes less than 1.5 meters in radius, the CRE age is allowed to increment within the model. Two TrackMet models are implemented to test both \edits{(3) Juno} and \edits{(6) Hebe} as parent sources. For \edits{(3) Juno}, particles are initiated at \edits{(3) Juno}'s semi-major axis and evolve according the seasonal and diurnal Yarkovsky effects. If a test particle reaches the 3:1J, 5:2J, or 8:3J resonances between 298 and 302 Myrs after the family formation event, it is removed from the model and tabulated for analysis; for (6) Hebe we remove particles encountering the $\nu$6 and 3:1J resonances between the same times. Particles found in the 8:3J resonance are also tabulated for analysis, but are allowed to continue evolving outward to the 5:2J to improve statistics. 
This process is repeated for 10 million initial particles. From the particles that reach resonances with timestamps between 298 and 302 Myrs, we then sample each according to their probability of Earth impact; 0.01 for the $\nu$6, 2.0$\times$10$^{-3}$ for the 3:1J, 2.0$\times$10$^{-4}$ for the 5:2J \citep{1997Sci...277..197G,nesvorny_asteroidal_2009}, and 1.5$\times$10$^{-4}$ for the 8:3J \citep{1997Sci...277..197G,de2010origin}\footnote{We independently tested the impact probability for particles entering the 8:3J with an inclination similar to \edits{(3) Juno} via a Rebound simulation of all of the major planets and 100 test particles that we evolved for 20 Myrs. All instances where particles achieved Tisserand parameters with the Earth less than 3 were identified and collision probabilities calculated according the collision probability defined by an {\"O}pik-algorithm based code \citep{opik_collision_1951,wetherill_collisions_1967}. We found that over 20 Myrs, the probability of a particle colliding with the Earth was 7.6$\times$10$^{-4}$. Given that we are only removing particles between 298 and 302 Myrs in the TrackMet simulation we thus use a collision probability of 1.5$\times$10$^{-4}$.}.  This allows us an independent method to determine the likely CRE ages of \edits{(3) Juno}-derived members reaching near-Earth orbits. The lack of collisional evolution in the Rebound model makes it necessary to explore any discrepancy between TrackMet and Rebound and evaluate the limitations of each model. Such a discrepancy could indicate either a) that a specific secondary collision event is required for \edits{(3) Juno} to be an adequate H chondrite source or b) \edits{(3) Juno} is unlikely to be a significant source of H chondrites with the current observed properties. 

\section{Results}\label{sec:Results}
\subsection{Rebound Results}

\begin{figure}
    \centering
    \includegraphics[width=0.7\linewidth]{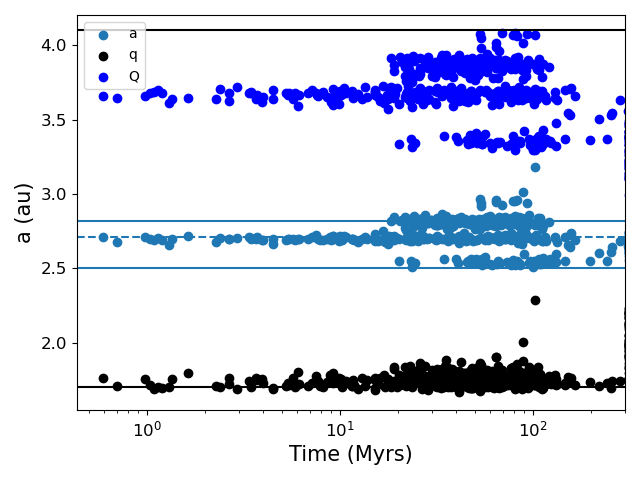}
    \caption{Semi-major axis (a), perihelion (q), and aphelion (Q) distances of particles that were removed from the simulation via intersections with Jupiter or Mars (top and bottom horizontal black lines). The 5:2 and 3:1 resonances with Jupiter are marked with the upper and lower horizontal blue lines, while the 8:3 with Jupiter is marked with a dashed blue line. Notice that the vast majority of particles were removed via interactions with Mars. Of similar interest is the apparent effectiveness of the 8:3 at scattering particles: of the 465 scattered particles, 131 had $a$ within 0.1 au of the resonance. }
    \label{fig:scatter_props}
\end{figure}

\begin{figure}
    \centering
    \includegraphics[width=0.7\linewidth]{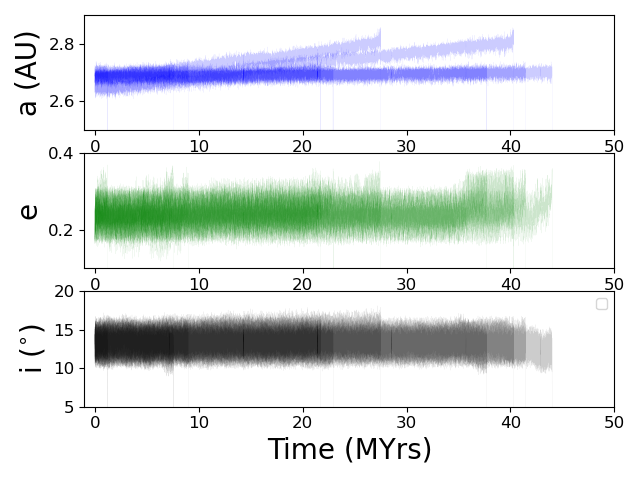}
    \caption{Evolution of semi-major axis, eccentricity, and inclination for 10 particles randomly selected from the scattered particles in the simulation. This sample all scatters within the first 50 Myrs. }
    \label{fig:particle_histories}
\end{figure}

A few representative particle histories from the Rebound simulations are shown in Figures \ref{fig:scatter_props} and \ref{fig:particle_histories}, respectively. As expected, the smallest particles, those between 0.5 and 10 meters in size, quickly evolve to the 5:2J and 3:1J resonances before scattering, typically within 100 Myrs (Fig. \ref{fig:particle_histories}). This evolution makes it clear that the original small members of the \edits{synthetic} Juno family-forming collision $\sim$300 Myrs ago have been effectively cleared out, with just 14\% of test particles smaller than 10 m remaining,  and would no longer be a significant source of H chondrite meteorites. These smaller particles also have collisional lifetimes of less than 30 Myrs \citep{bottke_linking_2005}, far less than the total simulation time of 320 Myrs. Larger particles in the simulation, with less significant acceleration due to the Yarkovsky force, take longer to migrate into the resonances. By the end of the simulation we find that 88\% of the initial 217 particles greater than 1 km in diameter are still in main belt orbits, and it is these sized objects that would have collisional lifetimes on the same timescale as the simulation. We turn our attention to these larger test particles.  

\begin{figure}
    \centering
    \includegraphics[width=0.7\linewidth]{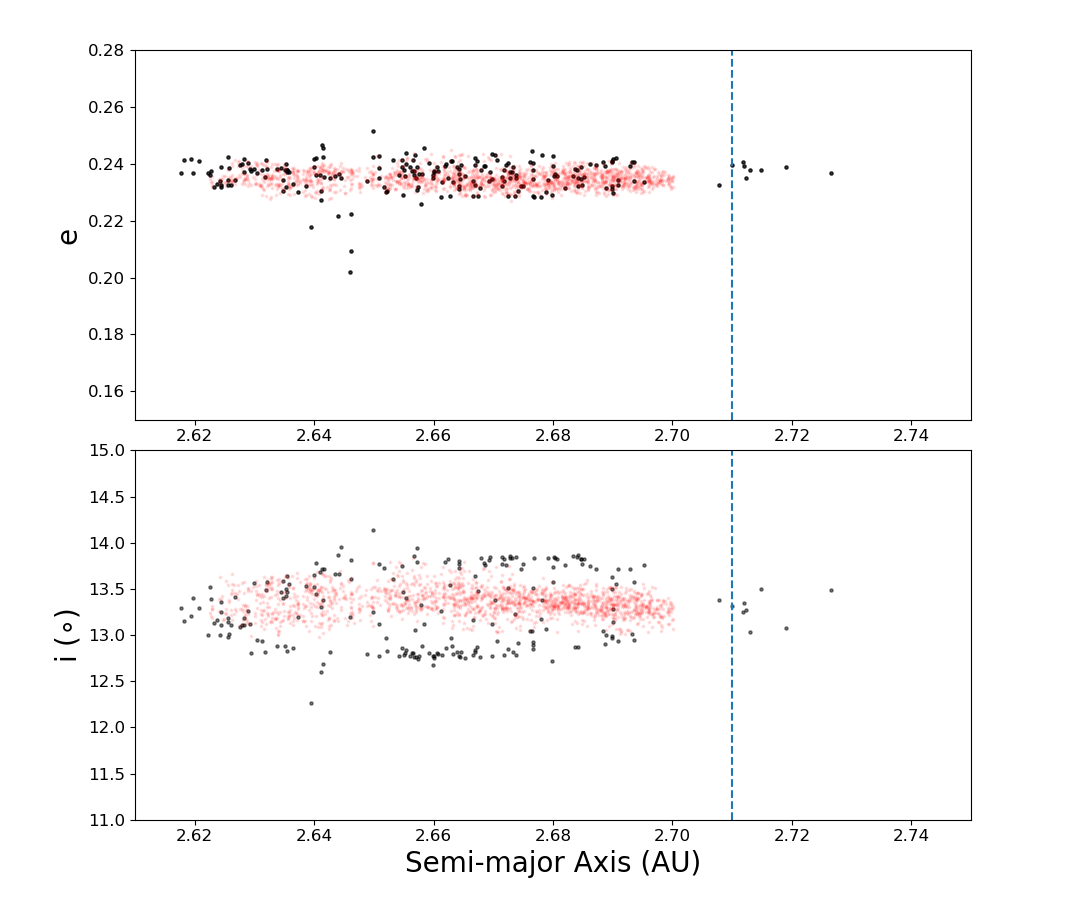}
    \caption{Comparison of the proper elements of the remaining test particles larger than 670 m in diameter in the Rebound simulation (black) with the proper elements of \edits{observed} Juno family members (red). \edits{First order proper} elements for the synthetic Juno family members have been determined via an average of the last \edits{100,000} years from the Rebound simulation archive, while the proper elements for the \edits{observed} Juno family members are taken from the AstDyS web portal. \edits{We find an average $a$,$e$, and $i$ of 2.660$\pm$0.024 au, 0.236$\pm$0.006, and 13.27$\pm$0.40$^{\circ}$ for the synthetic family, in agreement with the values of 2.667$\pm$0.021 au, 0.235$\pm$0.003, and 13.36$\pm$0.14$^{\circ}$ from the AstDyS database. }}
    \label{fig:juno_family_comp}
\end{figure}

A key test to underscore the validity of these simulations of the \edits{synthetic} Juno family evolution is to compare the final large test particles to \edits{observed} Juno family members (Figure \ref{fig:juno_family_comp}). 
We do this comparison in proper orbital semimajor axis, eccentricity, and inclination.
We use the list of \edits{observed} Juno family members and their synthetic proper elements reported in the AstDyS database's family lists\footnote{see \url{https://newton.spacedys.com/astdys/index.php?pc=5}} \citep{Knezevic:2003,Milani:2014,Knezevic:2014}.
For the Rebound particles, we approximate the proper elements by taking the average $a$,$e$, and $i$ from the last \edits{100,000} years of the simulation for \edits{each particle. While this method would not be accurate enough to determine family members from a larger main belt population, we find it is satisfactory for comparison to the AstDyS proper elements derived using a Fast Fourier Transform technique.}
The final semi-major axes of the D $\geq$ 670 m test particles in the Rebound simulations are in good agreement with the proper elements of the \edits{observed} Juno family members \edits{(see Fig. \ref{fig:juno_family_comp} caption)}, all of which are larger than 670 m in diameter, assuming that their albedo is similar to \edits{(3) Juno}'s (0.238). However, the spread of inclinations of the particles in our simulation is a little larger, which \edits{is likely the result of our uniform distribution of initial velocity vectors relative to the surface of (3) Juno. Real collisions would not likely produce such a uniform debris cloud.}.  These are the family members that will have collisional lifetimes of the correct scale to experience disruption near the end of the 300 Myr family age and are most likely responsible for the contributing Earth-crossing particles with CRE ages less than 20 Myrs in the complimentary TrackMet model \citep{farinella_meteorite_1998,bottke_linking_2005}. Given the model input for our synthetic family members from our simplified collision setup we are pleased with the similarity between the observed and modeled Juno family members.

\subsection{TrackMet Results}
\begin{figure}
    \centering
    \includegraphics[width=0.7\linewidth]{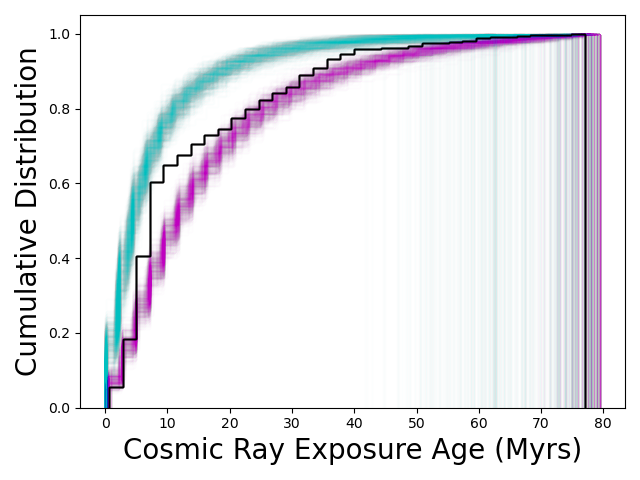}
    \caption{CRE age cumulative distributions for 500 resamplings of all Earth impacting particles originating from \edits{(3) Juno} (magenta) and \edits{(6) Hebe} (cyan) in the TrackMet model, with the measured H chondrite sample from \cite{marti1992cosmic} on Earth overlaid in black. The breakdown of impactor frequency for each resonance for \edits{(6) Hebe} and \edits{(3) Juno} is shown in Figure \ref{fig:res_breakdown}. Notice that the average cumulative distribution for \edits{(6) Hebe} overpredicts the number of H chondrites with CRE ages under 20 Myrs. \edits{(3) Juno} appears to match the measured distribution between 0 -- 5 Myrs and 20 -- 80 Myrs, but is depleted relative to the measured between 8 -- 15 Myrs.  }
    \label{fig:CRE_distributions}
\end{figure}

The CRE ages produced by the TrackMet model provide an additional line of support for the H chondrites originating from \edits{(3) Juno}. The 10 million initial particles in the TrackMet simulation are evolved collisionally and radially, and those that reach the 3:1J, 5:2J, 8:3J, and $\nu$6 resonances  are randomly sampled 500 times with terrestrial impact probabilities of 2.0$\times 10^{-3}$,  2.0$\times10^{-4}$,1.5$\times10^{-4}$ and 1.0$\times 10^{-2}$, respectively, to produce averaged CRE distributions. The TrackMet code handles the terrestrial impact probabilities for the 8:3J post-simulation to allow particles evolving outwards to reach the 5:2J. The CRE age distribution of these impacts for \edits{(3) Juno} is shown in the top panel of Figure \ref{fig:CRE_distributions} plotted with the actual H chondrite distribution from \cite{marti1992cosmic}.  The CRE age distribution was also modeled for family members originating from (6) Hebe, another potential candidate as a parent body for the H chondrites \citep{gaffey_asteroid_1998, vokrouhlicky_efficient_2000,fieber2020family}, using the $\nu$6 and 3:1J resonance and the corresponding impact probability of 1$\times 10^{-2}$ for the secular $\nu$6 resonance. Due to the proximity to those resonances only 1 million initial particles are needed to obtain good statistics on particles removed between 298 and 302 Myrs. The particle creation and Yarkovsky effect are implemented identically, but the particles are given \edits{(6) Hebe}'s orbital parameters to start and removed when they reach the $\nu$6 and 3:1J. These two simulations allow us to directly model the expected CRE age distributions of ancient family forming impacts from \edits{(3) Juno} and \edits{(6) Hebe} and compare them to the measured H chondrite CRE age distribution. 

\begin{figure}
    \centering
    \includegraphics[width=0.7\linewidth]{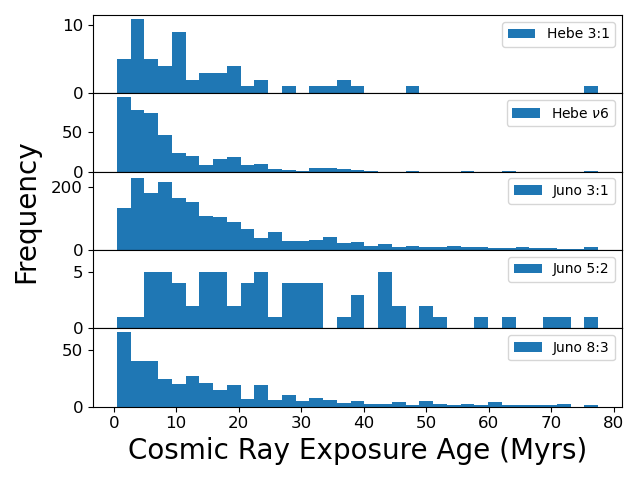}
    \caption{Contribution of each resonance to the delivery of Earth-impacting particles for \edits{(3) Juno} and \edits{(6) Hebe} TrackMet models. Note that the \edits{(6) Hebe} CRE distribution is skewed towards younger CRE ages due to both \edits{(6) Hebe}'s proximity to and the high efficiency of the $\nu$6 resonance .}
    \label{fig:res_breakdown}
\end{figure}

Figure \ref{fig:CRE_distributions} shows that the observed H chondrite distribution is difficult to match directly. The \edits{(3) Juno} distribution has a broad peak that is consistent with the H chondrite distribution, albeit lacking the steep peak at 7-8 Myrs. However, the TrackMet model makes it clear that the \edits{(6) Hebe} distribution is a poor match, with a CRE peak near 3 Myrs, rather than the 7-8 Myrs seen in the measured H chondrite distribution;  the \edits{(6) Hebe} model over predicts the number of H chondrites that would be expected to have CRE ages under 20 Myrs. The source of the discrepancy can be traced to two factors: distance from the parent asteroids to the nearest resonance and the associated efficiency of that resonance for delivering earth impactors (Fig. \ref{fig:res_breakdown}. The order-of-magnitude higher efficiency of the $\nu$6 resonance compared to the 3:1J, in addition to \edits{(6) Hebe}'s close proximity to both, means that a large number of family members can be delivered to Earth-crossing orbits before achieving $>$5 Myrs of cosmic ray exposure. In comparison, \edits{synthetic} Juno family members reach the 3:1 resonance with approximately the right peak CRE age, albeit with a much broader distribution. The 8:3J \edits{contributes} a significant portion of low CRE age objects, but the 5:2J only contributes a relatively small portion with a wide range of CRE ages. The sharp peak in the measured H chondrite CRE age distribution is not easily recognized as the result of one particular resonance. Our interpretation of this will be discussed further in Section \ref{sec:Discussion} in the context of family forming collisions in the main belt. 

\begin{figure}
    \centering
    \includegraphics[width=0.7\linewidth]{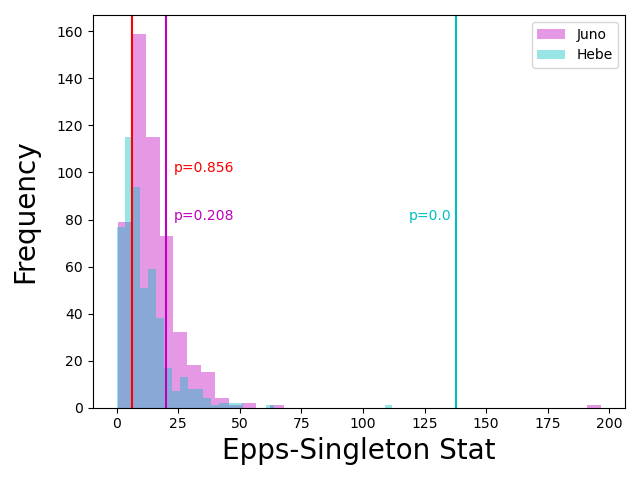}
    \caption{ Epps-Singleton statistical test results for 500 subsamples of 431 particles from the modeled \edits{(3) Juno} and \edits{(6) Hebe} TrackMet distributions compared to their total CRE distribution. ES tests were run for each re-sampling of the TrackMet distribution to understand the likelihood of producing the known H chondrite CRE age distribution from the model. The ES values resulting from the comparison of the measured H chondrite CRE age distribution with the total \edits{(3) Juno}- and \edits{(6) Hebe}-sourced distributions are shown as solid magenta and cyan vertical lines, with their bootstrap $p$-values labeled. A low ES value (or high $p$-value) is consistent with being unable to reject the null hypothesis, that the two samples are drawn from the same population. The CRE ages resulting from a family forming impact on \edits{(6) Hebe} 300 Myrs ago is statistically distinguishable from the H chondrite CRE age distribution, while for \edits{(3) Juno} the answer is not quite as clear. The red vertical line is the ES value for comparing an H chondrite CRE age distribution that has had the 6-8 Myr peak ``trimmed" to test how the removal of a recent stochastic event's contribution changes the bootstrap $p$-value. Removal of the peak increases the $p$-value by just under a factor of 4, making distinguishing between the \edits{(3) Juno}-derived and measured H chondrite CRE age distributions much more difficult.}
    \label{fig:juno_stats}
\end{figure}

To quantify the similarity between the \edits{(3) Juno} and \edits{(6) Hebe} TrackMet modeled CRE distributions and the actual H chondrite distribution we performed the Epps-Singleton (ES) statistical tests on randomly selected particles using the SciPy \textit{stats} module. The Epps-Singleton test is designed to compare two populations without assuming that they are drawn from a continuous distribution, making it useful for this experiment \citep{epps1986omnibus}. For the ES test, the null hypothesis is that the two samples are drawn from the same distribution; high ES and low $p$ values indicate that the null hypothesis can be rejected, and that the samples are drawn from different populations. 
We first calculate the ES value for the 431 known H chondrite CREs \citep{marti1992cosmic} compared to the entire simulated sets of Earth-crossing particles from either the \edits{(3) Juno} or the \edits{(6) Hebe} TrackMet models.
We then bootstrap the $p$-values for these ES values by comparing subsamples of the TrackMet models to themselves.
We randomly sample 431 particles from each TrackMet dataset, then calculate the ES value of that subsample compared to the overall dataset we know they were drawn from. 
This resampling was done 500 times, with the ES tests carried out for each resampled population, this generates the expected distribution of ES values for each modeled population to determine the likelihood of the measured H chondrite CRE ages being drawn from the same distribution as the \edits{(3) Juno}/\edits{(6) Hebe} CRE ages. The frequency of ES stat and bootstrapped $p$-values derived from these tests are shown in Figure \ref{fig:juno_stats}. 

When randomly sampling 431 particles from the \edits{(6) Hebe} TrackMet model and comparing to the full Trackmet model, the ES test produced ES values between 0 and 115, with the majority of values smaller than 50. When comparing the measured H chondrite CRE age distribution to the \edits{(6) Hebe} model we find a value of 137.78, higher than 100\% of the resampled tests. This yields a bootstrapped $p$-value of 0, indicating that with high statistical significance a 300 Myr-old family forming event on \edits{(6) Hebe} cannot reproduce the observed H chondrite CRE distribution. The same cannot be said for statistical tests of the \edits{(3) Juno} TrackMet model (Figure \ref{fig:juno_stats}). The values derived from a similar resampling of the \edits{(3) Juno} dataset show a broader distribution of ES statistical values, with the ES stat limited to between 0 and 200, with the majority lower than 75. This places the ES stat calculated for the measured H chondrite distribution (20.2) smaller than 20.8\% of the 500 \edits{(3) Juno} subsamples, for a bootstrap $p$ value of 0.208. We cannot reject the null hypothesis that the measured H chondrite CRE ages are drawn from the same \edits{(3) Juno} resample distribution (higher $p$-value , lower ES stat average). As noted above, our model lacks a distinct peak between 6-8 Myrs, but the data is not statistically inconsistent with the model. We also tested the effects of removing the CRE age peak between 6-8 Myrs, which may be the result of a family member disruption, by randomly selecting only one third of the measurements between 6 and 9 Myrs. The resulting ES stat of 6.75 is smaller than 85.6\% of the resampling comparisons, for a bootstrapped $p$-value of 0.856, which improves the match between the measured H chondrite CRE ages and those modeled from a family forming impact on \edits{(3) Juno} 300 Myrs \edits{ago}. We note that the Kolmogorov-Smirnov test, which assumes a continuous distribution, produces similar $p$-values for both the \edits{(3) Juno} and \edits{(6) Hebe} comparisons to the H chondrite CRE age distribution. We prefer to use the ES test since it does not require the continuous distribution assumption.  While this is far from definitive, it is worth noting that the CDF of the \edits{(3) Juno} TrackMet model appears to be a much better fit at small and large CRE ages compared to \edits{(6) Hebe}. The peak at 6-8 Myrs in the H chondrite CRE age distribution continues to foil attempts to model the distribution. 


\section{Discussion}\label{sec:Discussion}
The combination of the Rebound and TrackMet models suggests that \edits{(3) Juno}'s early efficiency at directly delivering primary collisional fragments to Earth-crossing orbits may be substantially supplemented by the flux of secondary objects resulting from the collisional evolution within the family. In this manner a 300 Myr old family forming event can continue to produce a meteorite population with a relatively young CRE age distribution. Given the similarity of the measured CRE distribution of H chondrites and the model, we now turn to examine the dynamical implications of \edits{(3) Juno} as an H chondrite source as well as future studies to test the hypothesis. 

One interesting component of the Rebound modeling is the importance of the 8:3J mean motion resonance in removing \edits{(3) Juno} family members in the first 100 Myrs of the simulation. As shown in Figure \ref{fig:scatter_props}, in the first 20 Myrs nearly all scattered particles are removed via interactions with Mars while in the 8:3J; this is consistent with the 8:3J being the first MMR that the test particles reach. Between 20 Myrs and 100 Myrs, we see the most interaction with all three resonances, and it is during this period that most of the scattered particles are removed from the Main Belt. In the last 100 Myrs of the simulation only 6 of the 1000 initial particles are scattered, and those are all the result of interactions with Mars. 
These last six particles are of particular interest because they represent objects that could still be on Earth-crossing orbits in the present day. 

These last scattered particles are between 0.5 and 500 m in diameter, and objects larger than 300 m have a collisional lifetime on the order of their scattering lifetimes in the main belt\citep{bottke2005linking}. While four of the six particles appear to have entered the 3:1J, indicating that their dynamical half-life is less than 3 Myrs in the inner solar system \citep{1997Sci...277..197G}, the most recently scattered particles are both injected into the inner solar system via the 8:3J. Unfortunately these are also small family members (D$<$30 m) that likely would have been disrupted within 100 Myrs of formation by the collisional evolution that is not modeled in our Rebound simulations. We note that at the end of our simulation we find $\sim$20 \edits{(3) Juno} family members very near the 8:3J resonance (Figure \ref{fig:juno_family_comp}). If these simulated family members are proportional to \edits{(3) Juno}'s current family, then we could expect $\sim$9\% of \edits{(3) Juno}'s \edits{observed} family population to have recently been within the 8:3J. Objects entering the inner solar system via the 8:3J have dynamical half-lives that are approximately an order of magnitude larger than those entering via the 3:1J, making it possible that large (D$>$300 m) Juno family members that have arrived to the 8:3J in the last 30 Myrs may still be present \citep{1997Sci...277..197G}. From that paper, and our own Rebound analysis of the 8:3J, we find that for these objects the time to enter Earth-crossing orbits is 11 Myrs on average. 
Given the number of \edits{observed} Juno family members with D$>$670 m and our own fraction of \edits{(3) Juno}-derived particles of similar size that have either been scattered from or recently arrived to the 8:3J in the last 30 Myrs of the simulation, we find that the presence of a large scattered Juno family member in the inner solar system should be rare in the present day. 
Assuming that 9\% of the \edits{observed} Juno family (1692 members) have passed through or are passing through the 8:3J in the last 30 Myrs, we can then calculate the number of R$>$335 m members remaining on an Earth-crossing orbit from the 8:3J as:
\begin{equation}
    N_{NEO,8:3} = N_{8:3J}p_{EC}e^{-\tau^{-1} T_{EC}} 
\end{equation}
where the number of Juno family members over 670 m in diameter entering the 8:3J is  N$_{8:3J}$ = 0.09$\times$1692 =152  , p$_{EC}$ = 7$\times$10$^{-8}$ yr$^{-1}$, $\tau_{EC}$=11 Myrs, and T = 30 Myrs. 
With these values we find that approximately 105 large (D$>$670 m) Juno family members have become Earth crossing in the last 30 Myrs and 7 can be expected to still be present in Earth crossing orbits, conservatively assuming they all entered 30 Myrs ago. This calculation depends linearly on the efficiency of Earth crossing orbits delivered from the 8:3J, so we will say that this is likely an optimistic calculation within an order of magnitude of the true value given the slow outward Yarkovsky drift of these larger family members. We note that this delivery mechanism is mentioned specifically for the \edits{8:3J} in \cite{1997Sci...277..197G}. 

This estimate is consistent with the recent discovery of shock-darkened asteroid (52768) 1998 \edits{OR$_{2}$} (D=2.5 km) on a near-Earth orbit \citep{battle2022shock}. While having a spectrum that is best classified as an Xn type in the Bus-DeMeo taxonomy \citep{demeo_extension_2009}, the NIR spectrum can be well fit with the reflectance spectrum of the shock-darkened H5 meteorite Chergach \citep{battle2022shock}. This discovery is encouraging, indicating that there may be up to 6 more Juno-family members in the inner solar system. We note that while the presence of an \edits{NEO} with an H chondrite-like spectrum is to be expected from a Juno-family forming event 300 Myrs ago, the same can not be said for a Hebe family member. Their dynamical half-life via injection into the 3:1J or $\nu$6 is much shorter (2.1-2.3 Myrs), making it much harder to explain large family members in near-Earth space without a larger flux of family members currently entering both resonances. Given the proximity of \edits{(6) Hebe} to both the 3:1J and $\nu$6, even at the slower expected radial drift rates due to the Yarkovsky effect \citep{bottke_yarkovsky_2006} one would expect that this flux in the last 10 Myrs would be minimal. Indeed the evidence that large Hebe family members are present on both sides of the 3:1J suggests as much \citep{fieber2020family}. 

This still leaves an open question regarding the difference in the cumulative distribution functions of CRE ages for the \edits{(3) Juno} model and the measured H chondrites: what is the source of the significant number of H chondrites that have a CRE age between 6-8 Myrs? While the TrackMet model is a good match to the younger ($<$6 Myr) and older ($>$8 Myr) H chondrite CRE distribution, the model is unable to explain the 6-8 Myr peak. No amount of resampling the \edits{(3) Juno} TrackMet distribution could reproduce this peak, so we must hypothesize about a possible origin. A spike in 6-8 Myr H chondrites would be consistent with a catastrophic breakup of a large Juno family member just outside the 3:1J, with 1-10 m size particles drifting inwards via the Yarkovsky effect reaching the 3:1J in 2-3 Myrs, and reaching Earth-crossing orbits within another 4-5 Myrs \edits{\citep{1997Sci...277..197G}}. This is not a particularly satisfying conclusion, as it requires an additional breakup event, but the collisional lifetimes of 1-2 km objects in the main belt outside of the 3:1J are on the order of 300 Myrs \edits{\citep{farinella_meteorite_1998,bottke_linking_2005}}. Of course, this CRE distribution spike might not be the result of a stochastic family member breakup near the 3:1J and could instead point to a different location in the main belt for the H chondrite parent body or indicate that another separate H chondrite parent is required to properly fit the measured distribution. 

One possible route to explain this is that specific spikes in the CRE age are sourced from smaller and more recent collisons on \edits{(6) Hebe}, while the background H chondrite CRE distribution is sourced from the family forming impact on \edits{(3) Juno}. \cite{marsset_3d_2017} identified five smaller craters on \edits{(6) Hebe} that could be the source of such material, and \cite{fieber2020family} linked these craters as potential sources specifically for the 6-8 Myr and 33 Myr CRE peaks in the H chondrite distribution\edits{.} Such a contribution would also help to explain the relatively even distribution of H chondrite falls with orbits linked to both the 3:1J and $\nu$6 \citep{borovicka2015small}. While it is possible for asteroids drifting radially to jump resonances, (i.e. Juno family members jumping the 3:1J and reaching the \edits{$\nu$6}), the efficiency of this process is not well established \citep{bottke2000dynamical}. This linear combination of distributions would likely be able to fit the measured one, but would introduce difficulty in explaining the widely accepted theory that the H chondrites are derived from a single major source based on isotopic and chemical evidence from the meteorite samples. This is not inconsistent with observations of the large H chondrite-like asteroids in the main belt, which appear to have all formed very rapidly and should be compositionally similar \citep{vernazza2014multiple}, but does still present a hurdle. Of these objects \edits{(3) Juno} is the largest, which would allow the widest range of thermal processing following accretion to occur without invoking alteration for fragments after a collisional event. \edits{(3) Juno}'s large size make it favorable to address the variety of thermally processed H chondrites that are in the meteoritic catalog, but \edits{(6) Hebe}'s size is only just below the limits imposed by paleomagnetic measurements \edits{(190 km vs. 230-320 km, \citealt{ruzicka_electron_2018})}. Clearly, future efforts to fit collisional events to observed CRE age distributions could provide more constraints on the many degeneracies of this problem, but also risk over-interpreting the available data.  

If the breakup of a Juno family member or separate impacts on \edits{(6) Hebe} are the source of this 6-8 Myr CRE age spike in the distribution, it may be possible to search for properties within that ``spike" population that are unique compared to the broader H chondrite sample. Ideally this would mean laboratory investigations of the H chondrite samples to search for any differences in 1 and 2 micron band depth absorption and center wavelength, yielding information about the relative abundances of olivine and pyroxene as well iron and/or calcium abundance within the minerals \citep{sanchez_composition_2015,noonan_search_2019,fieber2020family}. Such a re-analysis represents a substantial effort for NIR sample analysis, but could reveal promising clues about the origin of the H chondrite CRE age peak. Additional high resolution spectroscopy measurements of both \edits{(6) Hebe} and \edits{(3) Juno} to further determine differences in mineralogy are necessary to help guide new meteoritic re-analysis. \edits{The targets are eminently visible with both ground observatories, like the Legacy Survey of Space and Time planned for the Vera C. Rubin Observatory \edits{\citep{2023ApJS..266...22S}}, and space-based assets,  like the JWST, which can observe at new IR wavelengths \edits{\citep{2016PASP..128a8001M}}}. 

\section{Summary}\label{sec:Summary}
In this work we presented dynamical models of \edits{synthetic} Juno family members between 1 meter and 10 kilometers in radius. This was done to determine the viability of \edits{(3) Juno} as a dominant supply of H chondrite material here on Earth given the current properties and the constraints they impose on the H chondrite parent body. We find that if we assume the 300 Myr shock degassing age of most H chondrites is indeed representative of the Juno family forming event, several conclusions can be drawn:

\begin{enumerate}
    \item When only examining dynamical evolution, 14\% members of the \edits{synthetic} Juno family smaller than 10 meters in radius remain in main belt orbits after 300 Myrs, compared to 88\% of test particles larger than 1 km in diameter.  
    \item Small (D$<$10 m) primary family members of (3) Juno cannot directly supply the observed H chondrite flux; the most likely source would be larger Juno family members that underwent a significant, post-family forming event collision, as evidenced by the similarity in the family age and collisional lifetime of D$>$300 m members.  
    \item The collisional evolution of the \edits{synthetic} Juno family resulting from a family forming event 300 Myrs ago results in an Earth-crossing asteroid CRE age distribution that peaks at 6 Myrs with a significant tail at higher CRE ages.  A similar \edits{(6) Hebe} family forming event CRE age distribution peaks at 3 Myrs with a less significant large-age tail. 
    \item The current measured CRE age distribution of the H chondrite meteorite sample is statistically distinct from randomly drawn samples from the \edits{(6) Hebe} TrackMet model. The \edits{(3) Juno} TrackMet model appears to accurately represent a background H chondrite CRE age distribution. An additional source, likely a secondary breakup event 6-8 Myrs to go or discrete impacts on \edits{(6) Hebe}, is necessary to explain the peak in the measured H chondrite CRE age distribution if \edits{(3) Juno} is the parent body. 
    \item \edits{(6) Hebe} and \edits{(3) Juno} produce distinctly different CRE age distributions as parent bodies that are easily distinguished from one another.
    \item If \edits{(3) Juno} is a major parent body of the H chondrites, we can expect $\sim$7 asteroids in near-Earth orbits that are km-scale Juno family members at this time, consistent with the discovery of the shock-darkened asteroid (52768) 1998 \edits{OR$_{2}$} that exhibits a H chondrite like NIR spectrum. 
\end{enumerate}

 The feasibility of \edits{(3) Juno} as a parent body for the H chondrite relies heavily on the the efficiency of the 8:3J resonance to create NEOs, but it is difficult to directly model the collisional and dynamical histories of test particles simultaneously over the 300 Myr timescale with a large enough sample size to obtain good statistics. Pursuit of a more efficient method to achieve this will refine our dynamical modeling and exploration of the possible pathways to explain the observed meteorite properties and confirm parent body asteroids. We encourage the measurement of cosmic ray exposure age for newly discovered meteorites to expand the sample size for forward models to work with. A multi-pronged approach to understanding the lineage of the meteorite-asteroid relationship incorporating compositional and dynamical information is necessary, and a method we hope see implemented in the future. 

KV acknowledges support from NASA (grants 80NSSC19K0785, 80NSSC21K0376, and 80NSSC22K0512) and NSF (grant AST-1824869). DN acknowledges support from NASA's Solar System Workings program.

\bibliographystyle{cas-model2-names}
\bibliography{h_chondrite_ads,h_chondrite_zotero,h_chondrite}
\end{document}